# Interlayer Dzyaloshinskii-Moriya interaction in synthetic ferrimagnets

Shen Li[1,2,3], Mouad Fattouhi[4], Tianxun Huang[5], Chen Lv[3,5], Mark C. H. de Jong[2], Pingzhi Li[2], Xiaoyang Lin[1,3,*], Felipe Garcia-Sanchez[4], Eduardo Martinez[4], Stéphane Mangin[5], Bert Koopmans[2], Weisheng Zhao[1,3,*] and Reinoud Lavrijsen[2]

[1] National Key Lab of Spintronics, International Innovation Institute, Beihang University, 311115, Hangzhou, China

[2] Department of Applied Physics, Eindhoven University of Technology, P.O. Box 513, 5600 MB Eindhoven, The Netherlands

[3] School of Integrated Circuit Science and Engineering, Beihang University, Beijing, 100191, China

[4] Department of Applied Physics, Universidad de Salamanca, Plaza de la Merced, 37008 Salamanca, Spain

[5] Institut Jean Lamour, UMR CNRS 7198, Université de Lorraine, Nancy 54011, France

*Correspondence: XYLin@buaa.edu.cn; weisheng.zhao@buaa.edu.cn

**The antisymmetric interlayer exchange interaction, i.e., interlayer Dzyaloshinskii–Moriya interaction (IL-DMI) has attracted significant interest since this long-range chiral spin interaction provides a new dimension for controlling spin textures and dynamics. However, the role of IL-DMI in the field induced and spin-orbit torque (SOT) induced switching of synthetic ferrimagnets (SFi) has not been uncovered. Here, we exploit interlayer chiral exchange bias fields in SFi to address both the sign and magnitude of the IL-DMI. Depending on the degree of imbalance between the two magnetic moments of the SFi, the amount of asymmetry, addressed via loop shifts of the hysteresis loops under an in-plane field reveals a unidirectional and chiral nature of the IL-DMI. The devices are then tested with SOT switching experiments and the process is examined via both transient state and steady state detection. In addition to field-free SOT switching, we find that the combination of IL-DMI and SOT give rise to multi-resistance states, which provides a possible direction for the future design of neuromorphic computing devices based on SOT. This work is a step towards characterizing and understanding the IL-DMI for spintronic applications.**

Synthetic antiferromagnets (SAFs) have gradually become one of the research focuses in the field of spintronics due to its combined advantages of both ferromagnet (high readability and writability) and antiferromagnet (high scalability, thermal stability and ultrafast magnetization dynamics)[1,2]. Consisting of two or more magnetic layers separated by spacers, SAFs have evolved rapidly towards various applications such as magnetic random access memory (MRAM)[3], racetrack[4], spin-torque oscillators (STO)[5], magnetic synapses[6], etc. To construct these SAFs, the symmetric interlayer exchange interaction (IEI), i.e., Ruderman–Kittel–Kasuya–Yosida (RKKY) interaction[7-9] is one important basis and widely utilized. The discovery of RKKY facilitated significant technological developments in the field of magnetic storage and spintronics[10-12]. Lately, its antisymmetric counterpart attracted numerous interest due to its potential to break the symmetry required for deterministic information writing using spin-orbit torque (SOT) and promote topologically non-trivial 3D spin textures such as hopfions[1,13,14]. Recently, researchers have verified the existence of such an antisymmetric IEI, classified as the interlayer Dzyaloshinskii–Moriya interaction (IL-DMI), theoretically[15] and experimentally[16,17]. Further, research has shown that IL-DMI can be used to enable many useful aspects in spintronic applications, such as the mentioned in-plane (IP) symmetry breaking to achieve field-free SOT switching[18,19], creating 360° magnetic domain wall rings[20] and asymmetric domain wall propagation[21,22].

However, for synthetic magnetic structures whose net magnetic moment is not completely canceled, that is, synthetic ferrimagnet (SFi), the role of IL-DMI in the field induced and SOT induced magnetization switching process has not been clearly unraveled. In this work, we present two different asymmetric magnetization switching behaviours of SFi with perpendicular magnetic anisotropy (PMA) under the action of IL-DMI and IP field. Based on the degree of imbalance between the two magnetic moments of SFi, the magnetization switching along the easy axis will appear either as two-step switching (see Figure 2a) or three-step switching (see Figure 2d) due to a high intrinsic PMA compared to the antiferromagnetic RKKY interaction. Depending on the additional IP field and the current magnetization configuration, IL-DMI can assist or hinder the magnetization switching processes, which changes the switching fields of the SFi. In order to make the results more intuitive, here, we use the macroscopic relative behavior of the branches in the hysteresis loop to describe the effect of the IL-DMI on the switching field. In two-step switching or three-step switching case with IL-DMI, the hysteresis loops will behave either as an overall shifting or expanding-shifting-contracting in three sequential switching branches, respectively. The outer loops of the three-step switching can be used to calibrate field misaligning errors to improve measurement accuracy. Moreover, the shifts of the inner loop indicating the switching between two antiparallel states will be greatly amplified as the effect of the IL-DMI on both layers switching needs to be considered. Both advantages of the self-calibration function and larger shift make the detection of IL-DMI more efficient in three-step switching. Progressing towards the influence of IL-DMI in SOT switching, we measure both the transient state of the SFi during the application of the current pulse and the steady state of the SFi after the current pulse with a time interval. In addition to field-free switching, we also find that the combination of IL-DMI and SOT can achieve multi-resistance state with an asymmetric critical switching current distribution, which provides a promising direction for the future design of SOT-based neuromorphic devices[23]. Corresponding micromagnetic simulations are in agreement with experimental results, further elucidating the role of IL-DMI in hysteresis loop and SOT switching of SFi. This work unraveled the distinct role of IL-DMI in its magnetization switching process and the corresponding application potentials.

## Asymmetric magnetization switching behaviours of SFi with IL-DMI

We start by developing the necessary concepts to clearly identify the effect of IL-DMI. In general, the magnetization reversal in SAFs is invariant under the inversion of the magnetic field direction. However, this field-reversal invariance is broken when there is IL-DMI in the SAF structure. Figure 1a shows the most prominent feature of IL-DMI during SAF magnetization switching: chiral exchange bias. The bias field $H_{IL\text{-}DMI}$ is opposite under two opposite antiparallel arrangements of the SAF. To measure such a variable-direction exchange bias, we apply a constant-direction IP magnetic field $H_{IN}$ during the hysteresis loop switching. Accordingly, the magnetization switching process will either be assisted or hindered depending on the relative direction of $H_{IN}$ and $H_{IL\text{-}DMI}$. Figure 1b and c show the measurement schematic of IL-DMI with anomalous Hall effect (AHE) under a constant field $H_{ext}$. For a Hall bar with each specific angle $\varphi$ in the X-Y plane, the magnetization switching process is measured by rotating the sample angle $\theta$ clockwise or counterclockwise in the X-Z plane, in which the direction of $H_{IN}$ will be reversed. During the rotation of $\theta$, the magnetic component perpendicular to the sample plane $H_Z = H_{ext} cos\theta$ drives the switching and the IP component $H_{IN} = H_{ext} sin\theta$ helps to define the IP switching direction. Further, we obtain the AHE loops at each angle $\varphi$ and compare them. The sign and magnitude of the DMI can be extracted from the asymmetric switching behavior driven by $H_{ext}$ with $H_{IN}$ along different directions.

Depending on the degree of imbalance between the two magnetic moments of the upper and lower layers, the SFi magnetization switching will appear as either two-step switching or three-step switching, respectively. When the magnetic moment difference between the two layers is small, the magnetization switching of SFi is similar to that of SAF. Each layer switches only once and it requires two steps to complete the switching. The IL-DMI is conventionally measured in a compensated SAF stack with such a two-step switching process[16,18,19,21,24-27]. Figure 1d shows the schematic of asymmetric switching of SFi in two-step switching case (corresponding to hystersis loop in Fig. 2a) due to IL-DMI and an additional IP field component. As demonstrated[16], in this all single-layer switching (SLS) situation, the hysteresis loops behave as a two-step switching. When $H_Z$ is ascending, every switching step is assisted since $H_{IL\text{-}DMI}$ and $H_{IN}$ are parallel and adding up. Similarly, when $H_Z$ is descending, the switching processes will both be hindered. This assist-assist-hinder-hinder behavior will hence appear as an overall shift in the full hysteresis loop. Therefore, the IL-DMI can be probed by measuring the overall loop shift (see Supplementary Section 1 for the analysis and simulation of the whole switching process) and the field strength of IL-DMI can be characterized by this shifting magnitude. Nevertheless, due to the small magnitude of IL-DMI, the shift is small and hard to extract relative to the underlying thermal stochastic processes randomizing the switching field[28]. Moreover, small errors in angle setting due to sample mounting and hysteresis in the rotation stage will also cause a small loop shift and this cannot be eliminated.

To improve upon the above issue, we propose to use the hysteresis loop of a SFi stack exhibiting a three-step switching process. Figure 1e shows the schematic of asymmetric switching of SFi due to IL-DMI when there is a magnetization rearranging process due to the imbalance of the magnetic moments of the two layers (corresponding to hystersis loop in Fig. 2d). When there is a large thickness difference between the upper and lower FM layers, an extra step in the hysteresis loop is observed as the thicker layer tends to align with the external magnetic field due to the Zeeman energy gain. However, this requires the other layer to also switch as the RKKY interaction is still strong enough to stabilize the antiferromagnetic state relative to applied field. We term such an event as double layer switching (DLS). In this three-step switching case, the switching sequence of the two FM layers is completely changed. The thinner layer will switch three times in all three steps and the thicker layer will only switch once in the second step. Accordingly, the hysteresis loop no longer exhibits as an overall shift as shown in the two-step switching of the SFi. During SLS in three-step switching, the effect of IL-DMI on FM is consistent with that described in two-step switching. However, the switching sequence changes of the two layers make the outer loops expand or contract instead of shifting. Here, the expansion (contraction) represents the change in the width of the outer loop, which occurs when the magnetization switching is both hindered (assisted) as $H_Z$ ascending and descending. In the process of DLS, since the effective field of IL-DMI will be reversed after the switching and the FM layers are antiferromagnetically coupled, $H_{IL-DMI}$ and $H_{IP}$ along the same direction will hinder its transition to the next state instead (see Supplementary Section 2 for the analysis and simulation of the whole switching process). Consequently, the three steps in the hysteresis loop will behave as expanding-shifting-contracting in sequence (for more information see Fig. S2d). Moreover, since the two layers are simultaneously switching, the effective IL-DMI energy difference will double in this process. Considering the exponential relationship between the energy difference and the switching field according to Arrhenius law[29], the switching field shifting degree in the inner loop will significantly increase. Therefore, we can utilize the hysteresis loop shift during DLS to measure IL-DMI more accurately. However, this conjecture implies that extracting quantitative energy scales for the IL-DMI ($J/m^2$) via linear Zeeman energy consideration (loops shift) to become questionable.

To demonstrate experimentally the aforementioned two types of asymmetric switching, we deposit SFi films of Ta(3)/Pt(3)/Co($t$)/Ir(1.45)/Co(1)/Cu(1)/Ta(3) (layer thicknesses in nanometers in parenthesise) by magnetron sputtering. See Methods for specific growth methods and the possible origin of IP symmetry breaking to generate IL-DMI. Cu is used to weaken the AHE of the capping layer, resulting in higher effective AHE signal difference in our stacks[30]. In order to verify the two types of hysteresis loop behavior, we fabricate three samples S_1, S_2, and S_3 (the numbers indicate the thickness difference in angstroms between the upper FM and the lower FM layers) by changing the thickness $t$ of the lower FM layer to 0.9, 0.8, and 0.7 nm, respectively. Accordingly, the hysteresis loop of S_1 behaves as a two-step switching and the hysteresis loops of S_2 and S_3 behave as three-step switching. AP+ and AP- represent the antiparallel states with magnetization up and down in the thicker upper Co layer, respectively. The basic magnetic properties and out-of-plane (OOP) AHE curves of the three samples are presented in Supplementary Section 3. Then the asymmetric AHE switching loops due to IL-DMI and an additionally applied $H_{IN}$ are measured using the configurations shown in Fig. 1b and c.

We first measure the AHE loop of S_1 to verify the consistency of IL-DMI behaviour in the SFi with two-step switching and the SAF with two-step switching reported before[16]. As shown in Fig. 2a, the AHE loops at the azimuth angle φ=135° and φ=315° are slightly shifted to left and right, respectively. The vertical dashed lines indicate the maximum AHE loop shift of SFi S_1 corresponding to the switching process shown in Fig. 1d. The switching field difference $\Delta\mu_0 H_{SW} = 2.7$ mT is obtained by averaging the two shifts for switching from parallel to antiparallel (shown by dashed lines and arrows). Theoretically, these two shifts should be equal. However, the shifts measured in actual experiments are generally different. The difference between the two shifts is inevitable, which is caused by an angle calibration error or misaligned magnetic field. Figure 2b and c show the azimuthal angular dependence of the switching field of the upper layer and lower layer of S_1 with the IP field, respectively. The magnitude of the IP field is kept as $H_{IN} = H_{ext} sin\theta$ as the direction of it rotates from 0 to 360°. We find that the magnetization switching of S_1 exhibits oppsite unidirectional anisotropy in the two FM layers. This asymmetric distribution of the switching field confirms the presence of IL-DMI in our sample and the asymmetry (AS) axis is along φ=145° and the symmetry (S) axis is along φ=55°. The switching field difference $\Delta\mu_0 H_{SW}$ is 2.8 (±0.7) and 2.7 (±0.5) mT for switching from antiparallel to parallel and from parallel to antiparallel, respectively.

With the basis of the experimental results of two-step switching, we now verify the proposed asymmetric switching behavior for three-step switching in a SFi. We measure the asymmetric AHE loops of sample S_2 to determine its IL-DMI bias field. As expected, in the case of three-step switching, the hysteresis loop behaves as the expanding-shifting-contracting under both $H_{IL-DMI}$ and $H_{IN}$. Figure 2d shows the maximum switching field shifts of SFi S_2 during the DLS process measured at azimuth angles of 112.5° and 292.5°, which corresponds to the switching process shown in Fig. 1e. At the azimuth angle φ=292.5° and φ=112.5° near to the AS axis, the inner loops for DLS are clearly shifted to left and right, respectively (see Supplementary Section 4 for the comparison between the maximum and minimum DLS shifts). Figure 2e shows the azimuthal angular dependence of the switching field during the DLS process of S_2. Figure 2f shows the azimuthal angular dependence of the switching field difference $|\Delta\mu_0 H_{SW}|$ between the switching of AP+ to AP- and AP- to AP+. We found the maximum switching field difference $\Delta\mu_0 H_{SW}$ is 22.1 (±3.4) mT. The expansion or contraction behavior of the outer loops are shown in Supplementary Section 5. Further, we confirm the magnitude of the IL-DMI field in S_2 to be 1.23 (±0.35) mT by calculating the switching field width difference of the outer loops in the SLS region, where the error bar represents the standard deviation. Theoretically, due to the expansion or contraction characteristics of the outer loops, we can calibrate the two hysteresis loops with the center of the outer loops as a reference to eliminate the shifting caused by angular errors or misaligned field. This procedure avoids the unequal situation of the shifts that appear in Fig. 2a. The deposition conditions of S_2 and S_1 are exactly the same, differing only in the thickness of the lower FM. Indeed, the intensity of the IL-DMI effective fields of the two (1.23 mT and 2.7 mT) is found to be close in magnitude. However, the loop shift of sample S_2 during DLS (22.1 mT) is one order of magnitude larger than the overall shift of S_1 (2.7 mT). We believe that this amplification in shift magnitude is caused by the simultaneous switching of both the upper and lower layers in DLS. According to Arrhenius Law[29], the doubled IL-DMI energy gap and the exponential change of the switching field with energy will significantly increase the field shift. With outer loops calibration and large-shifting in the inner loops, we discovered a more effective IL-DMI measurement in three-step switching process.

## Transient state detection in SOT switching

At present, one of the most promising applications of SFi materials is to utilize SOT to switch the SFi magnetic moment and apply it to data storage and logic computing[32]. For SOT switching, SFi has the advantages of high-speed, minimized magnetic crosstalk, extra robustness against external field perturbation, etc., and its net magnetic moment allows for easy magnetic reading[33,34]. However, there are currently only limited articles that have studied the SOT switching of SAF in the presence of IL-DMI. The research results are mainly focused on IL-DMI assisted SOT to achieve deterministic switching of the SAF with PMA[18,19]. After systematically studying the behavior of IL-DMI in the hysteresis loop switching, we further performed SOT experiments to explore the role of IL-DMI in the SOT switching process of our SFi devices. To comprehensively study the impact of IL-DMI on SOT switching, we apply two methods of detection: Method I, to detect the transient state right after the SOT switching pulse; Method II, to detect the steady state after the SOT pulse applying and a time interval. Figure 3a shows the schematic illustration of Method I. The current pulse width of SOT is 1 ms and the AHE voltage is picked up at the end of each $I_{pulse}$. In this case volatile and nonvolatile magnetization change induced by SOT are simultaneously probed. Figure 3b and c show the dependence of $R_{AHE}$ with $I_{pulse}$ in SFi S_3 and S_2 Hall bars (10-μm wide) with the assistance of different IP fields $H_X$. Under the joint action of IL-DMI, spin current and thermal effect, it is no longer the mutual switching of the two antiparallel states of AP+ and AP-[35]. The slow rising or falling behavior of $R_{AHE}$ before switching (range from -10 to 10 mA) indicates that the upper and lower layers of SFi tend to be arranged perpendicularly with the current increasing, which is caused by the IP polarization current of SOT, the chiral configuration preferred by IL-DMI, and thermal effects (see Fig. S5 for the similar behavior of $R_{AHE}$ under IP field). In previous work[35-38], the same method is used to detect SOT switching of SAF stacks without IL-DMI. To the best of our knowledge, none of these reported the multi-state switching curves. Therefore, we believe that IL-DMI plays a key role in this switching behavior. The red and blue lines represent the positive (+) and negative (-) current scanning results of $I_{pulse}$, respectively. First, no SOT switching occurs without $H_X$. Second, with proper $H_X$ assistance, deterministic SOT switching can be achieved.

When $H_X$ is reversed, the switching polarity of the SOT is also reversed. However, when $H_X$ exceeds a certain value (50 mT for S_3 and 100 mT for S_2), no SOT switching occurs and the curves of positive and negative current sweeping coincide. We believe this is because when $H_X$ is large enough, SFi will spontaneously switch to only one of the antiparallel states so that the chiral exchange bias field of IL-DMI is consistent with $H_X$. In this case, no matter the current is scanning positively or negatively, SFi prefers only one of the AP states and no deterministic switching will happen due to IL-DMI. It is also verified by seeing that the SFi only prefers the AP+ or AP- states when $H_X$ reaches +50 mT or -50 mT, respectively. One previous article reported that perpendicular magnetization switching of SAF with IL-DMI can be achieved using IP field only[31], which to some extent is consistent with our SOT results when $H_X$ reaches values larger than +50 mT or -50 mT. Further, we plot the dependence of the critical switching current of the device as a function of the IP field. As shown in Fig. 3d and e, regardless of the direction of $H_{IP}$, the absolute value of the critical current in the positive scan (red square) is always smaller than that in the negative scan (blue triangle) at each $H_X$ value. Similar results are also obtained in S_1 and the detailed switching results can be found in Supplementary Section 5. Overall, this asymmetric behavior occurs because $H_X$ and $H_{IL-DMI}$ always switch simultaneously. The reversal of $H_X$ will cause the polarity switching of the SOT. Accordingly, the chiral bias field due to IL-DMI will be reversed in another AP state. Thus, the unidirectional asymmetric behavior of the critical current is due to the fact that $H_X$ and $H_{IL-DMI}$ are always oriented in the same direction once the scanning direction is fixed. This result can in turn be used as a proof of the chiral exchange bias produced by IL-DMI.

## Steady state detection in SOT switching

After the transient resistance states detection, we also use Method II to detect the steady state to complete the research of the behavior of IL-DMI in SOT switching. Figure 4a shows the schematic illustration of Method II. In order to eliminate the influence of thermal effects on the switching, after each SOT pulse, an interval of 3 seconds is waited, until we apply a very small AC current to detect $R_{AHE}$ of SFi using the lock-in technique. In this case only the nonvolatile magnetization change induced by SOT is probed. Correspondingly, at this time the SOT switching curve turned into the conventional two AP state switching and the phenomenon of transient multistate no longer appeared. We use this method to detect the effective field generated by IL-DMI and verify whether it can assist SOT to achieve field-free switching. Here, we conduct experiments on symmetric Hall bars, where both stripes have the same width of 2 μm so that we can obtain the results with IP field along both φ=0° and φ=90° in one device. In this case, after measuring the SOT switching experiment with IP field along φ=0°, we can rotate the device and switch the current and voltage line at the same time to obtain the SOT switching result with IP field along φ=90°. Figure 4b and c show the SOT switching curve of S_2 with the Hall bar (2-μm wide) angle at φ=0° and φ=90° under different IP fields. In the case of φ=0° (φ=90°), almost no switching signal is detected in the sample when the IP field is -0.2 mT (0.1 mT) and the SOT curves show the opposite switching polarity before and after this field. These results indicate that the effective fields produced by IL-DMI at φ=0° and φ=90° are 0.2 mT and -0.1mT, respectively. We further demonstrate this effective field by plotting the AHE resistance difference before and after switching as a function of magnetic field shown in Fig. 4d. The above results show that the IL-DMI in our device can assist SOT to achieve field-free switching. The effective field changes the direction at 0° and 90°, which is consistent with the effective field direction of the IL-DMI. However, the magnitude of the effective field we measured is relatively small compared to the $H_{IL\text{-}DMI}$ measured in asymmetric hysteresis loop switching and this requires further investigation.

## Micromagnetic simulations for SOT switching

In order to explain these phenomena caused by the combination of IL-DMI and SOT, we proceed to use micromagnetic simulation to verify the present experimental results. See Method for the simulation details. Since the field-free SOT switching has been demonstrated in previous articles[18,19,39-41], we focus on the asymmetric critical current distribution and the multi-state formation before the switching. Figure 5a and b show the magnetization switching of SFi with SOT current with the assistance of positive and negative field, respectively. As the direction of IP field reverses, the SOT switching polarity also reverses, as shown by the rotation direction of the arrows in Fig. 5a and Fig. 5b. However, the critical current in the negative scan is always smaller than that in the positive scan. Figure 5c shows the simulation result of unidirectional asymmetric critical current distribution, which is consistent with the experimental results shown in Fig. 3d and e. The inset explains that this unidirectional asymmetric behavior is due to the SOT polarity switching as $H_{IN}$ reverses. Correspondingly, the effective IL-DMI field switches, too. Therefore, during negative current sweeping, $H_{IN}$ and $H_{IL\text{-}DMI}$ are always in the same direction and adding up. During the positive current sweeping, $H_{IN}$ and $H_{IL\text{-}DMI}$ are always in the opposite direction and subtracting. Therefore, the critical current in negative scan will always be smaller than that in positive scan.

For simulation results of multi-state formation before the SOT switching, see Supplementary Section 6 for details. We believe that the transient state generated in the experiment is the result of the perpendicular alignment of the upper and lower FM layer magnetic moments under the combined action of IL-DMI, SOT and thermal effects. We also need to consider magnetic domain nucleation and domain wall motion in μm-scale Hall bar devices. Given that our simulation size is 1000×250 nm and MuMax's simulation environment is at 0 K, it does not actually reproduce the magnetic domain wall motion and thermal effects. Figure S9a shows the SOT switching dynamics of SFi with different magnitude of IL-DMI. Figure S9b shows that under the action of IL-DMI, regardless of whether the SOT current reaches the critical value, the magnetic moments of the upper and lower layers will gradually be aligned vertically during the current applying. This can shed some light to the cause of the multi-resistance state formed before switching of SFi in our experiment.

## Conclusions

In conclusion, we unraveled the role of IL-DMI in the field induced and SOT induced magnetization switching of SFi structure. Under the combination of the IP magnetic field and IL-DMI, the hysteresis loop of SFi manifests as an overall loop shift and loop expanding-shifting-contracting in the case of two-step switching and three-step switching, respectively. In three-step switching, the center of the outer loops can be used for calibration and eliminate field or angle misaligning errors and the shift of the inner loop will be greatly amplified due to DLS. Both advantages make the detection of IL-DMI in three-step switching more efficient. We used both transient-state and steady-state resistance detection to explore the performance of IL-DMI in SFi's SOT switching. In addition to assisting SOT to achieve field-free switching, IL-DMI can also assist SOT to realize multi-resistance state before switching with asymmetric unidirectional critical current distribution. This provides a promising direction for future SOT-based neuromorphic computing research.

## Methods

**Sample growth, device fabrication and electrical measurement:** A series of Ta(3)/Pt(3)/Co(t)/Ir(1.45)/Co(1)/Cu(1)/Ta(3) (layer thicknesses in nanometers) stacks were deposited on thermally oxidized silicon substrates by DC sputtering under a base pressure lower than $10^{-8}$ Torr. We do not impose a special way of IP symmetry breaking on purpose. However, we speculate that the IP symmetry breaking is caused by the oblique sputtering of Co and Ir targets. One of the evidence is that the asymmetric axes of the IL-DMI in all our devices are along the second and fourth quadrants. In addition, some articles reported that IL-DMI can still be detected even in Co/Ir/Co control sample[31]. The origin of IL-DMI in our structure will be further elucidated in follow-up research. A superconducting quantum interference device (SQUID) was used to quantify the magnetic moments of each sample. The films were fabricated into Hall bar devices of 20, 10, 4 and 2 μm width by optical lithography and lift-off techniques. The measurement of IL-DMI is achieved through an MR setup. We apply a constant magnetic field first. By rotating the sample clockwise or counterclockwise, we can get the AHE hysteresis loop while applying a constant-direction IP field to the sample. The transient state and steady state detections in SOT measurement are achieved through Keithley 6221, 2182 and Keithley 33250, 6221, Stanford SR830, respectively.

**Micromagnetic simulation:** Micromagnetic simulations were carried out by solving numerically the LLG equation augmented with damping-like SOT that considers the contribution of the current flowing through the heavy metal layer. We included the contribution of the interlayer DMI as an effective field in the micromagnetic simulator Mumax3[42] with the energy form as $E_{DM} = D \cdot (S_1 \times S_2)$. The parameters of ferromagnetic material for the simulations are chosen to be $A_{ex}$ = 19 pJ/m (exchange stiffness)[43]; $M_S$ = 0.55 MA/m (upper layer) and 0.85 MA/m (lower layer) (saturation magnetization); $K_u$ = 0.26 MJ/m$^3$ and 0.52 MJ/m$^3$ (uniaxial anisotropy). The interlayer DMI and RKKY coupling strength are set as $D_{1,2}$ = 0.12 mJ/m$^2$ and $J_{RKKY}$ = 0.3 mJ/m$^2$, respectively. A typical heavy metal spin Hall angle of 0.3 is used[44]. The spin polarization is set as [0,1,0] when the current is along [1,0,0].

ACKNOWLEDGMENT

This work was supported in part by the National Natural Science Foundation of China (No. 62371019, T2394475, 92164206, and 52261145694), the Beijing Natural Science Foundation (No.4232070), the International Mobility Project (No. B16001) and the China Scholarship Council (CSC).

AUTHOR CONTRIBUTIONS

S.L. conceived the original idea. S.L., T.H., X.L., W.Z. and R.L. planned and designed the experiments. T.H. and C.L. fabricated the samples. S.L. performed IL-DMI measurements and steady SOT measurements. T.H. and S.L. performed transient SOT measurements. M.F., S.L., F.G.-S. and E.M. performed micromagnetic simulations. S.L. wrote the paper with M.C.H.J., P.L., X.L., S.M., B.K., W.Z. and R.L. All the authors participated in discussions of the research.

COMPETING FINANCIAL INTERESTS STATEMENT

The authors declare no competing financial interests.

FIGS

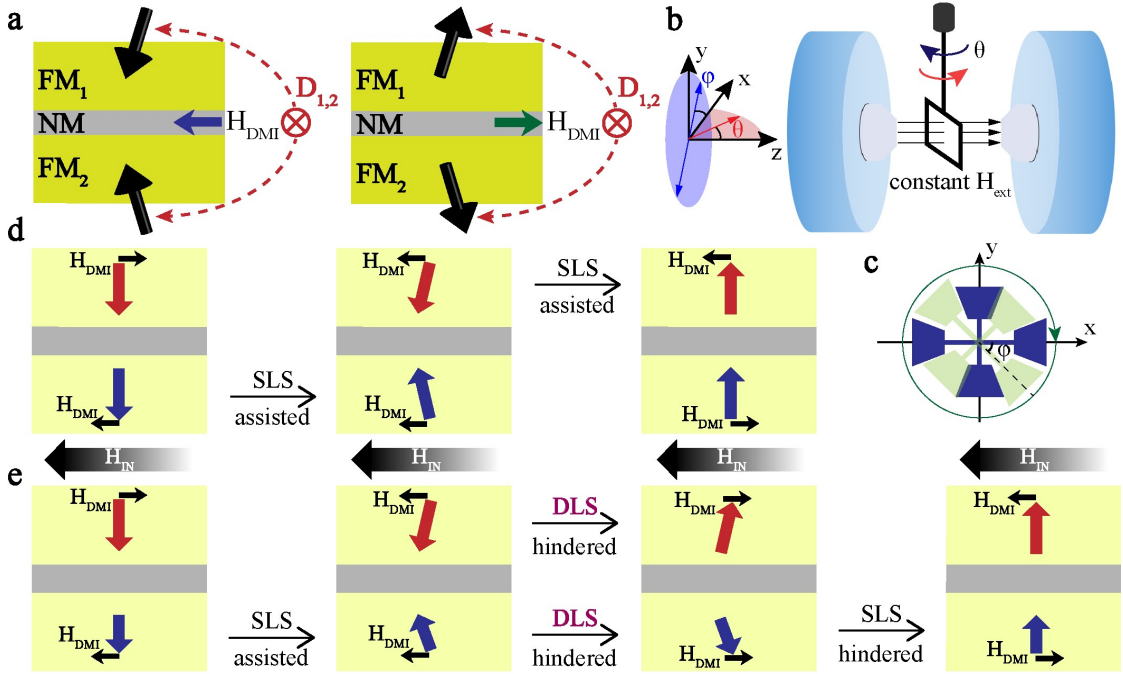

**Fig. 1 | Chiral exchange bias and field-reversal symmetry breaking of the SFi layer with IL-DMI. a**, The exchange bias field of the IL-DMI is opposite under two different antiparallel arrangements of the SAF with the same chirality. **b-c**, The measurement schematic of IL-DMI under a constant field. The field is set to a value which is slightly larger than the out-of-plane (OOP) saturation field. **b**, For each Hall bar with a specific angle $\varphi$ in the X-Y plane, the magnetization switching of the SFi is measured by rotating the sample angle $\theta$ in the X-Z plane, where $H_Z = H_{ext}cos\theta$ and $H_{IN} = H_{ext}sin\theta$. **c**, The Hall bar is located in the X-Y plane and rotates at an angle of $\varphi$ with the step of 22.5°. The angle rotating process in **b** is repeated for every different $\varphi$. **d-e**, Schematics of asymmetric switching of SFi due to IL-DMI and an additional IP field in two-step switching (**d**) and three-step switching (**e**) case. The red and blue arrows indicate magnetizations of the top and bottom FMs, respectively. The black arrows represent an

IP bias field. The chiral exchange bias $H_{IL-DMI}$ breaks the inversion symmetry between up-to-down (U–D) and down-to-up (D–U) switching polarities in the presence of $H_{IN}$. When the thickness difference between the upper FM and lower FM is large, the DLS occurs. Compared with single-layer switching (SLS), instead of assisting, $H_{IL-DMI}$ and $H_{IN}$ along the same direction will hinder the switching of both layers simultaneously.

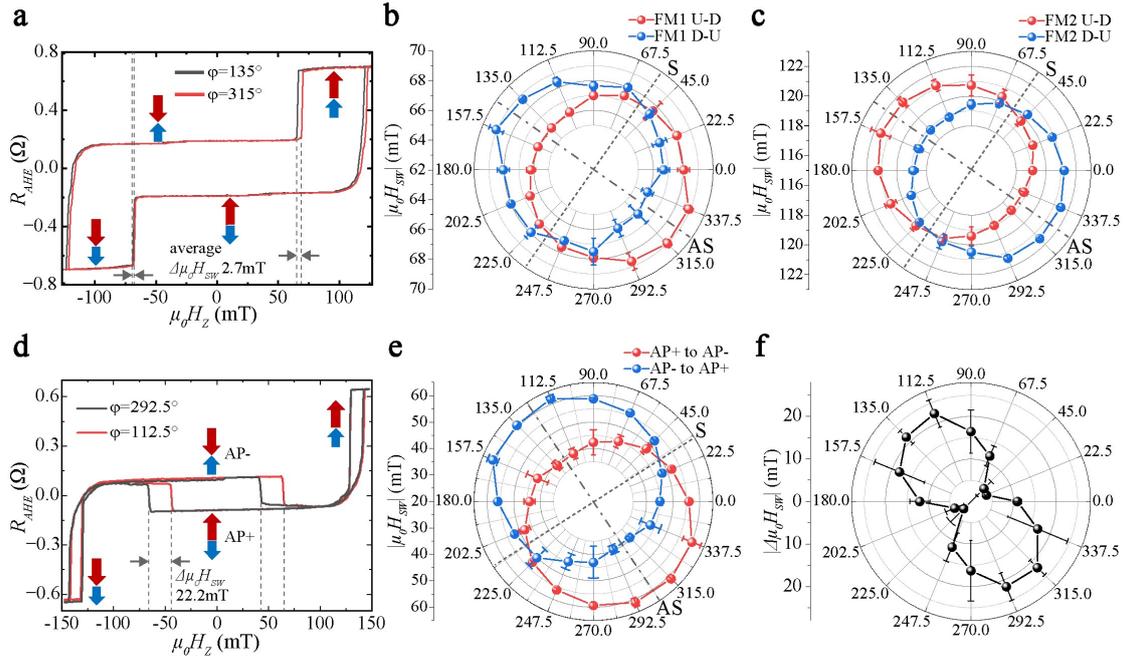

**Fig. 2 | Characterization of the IL-DMI effect for SFi in both two-step switching and three-step switching. a**, Maximum AHE loop shifts of SFi S_1 measured at azimuth angles of 135° and 315° in the case of two-step switching. **b-c,** Azimuthal angular dependence of the switching field of the upper layer (**b**) and lower layer (**c**) of SFi S_1. The asymmetric (AS) axis and symmetric (S) axis are along 145° and 55°, respectively. **d**, Maximum switching field shifts of SFi S_2 during the DLS process measured at azimuth angles of 112.5° and 292.5° in the case of three-step switching. **e**, Azimuthal angular dependence of the switching field during the DLS process of S_2. AP+ and AP- represent the antiparallel states with magnetization up and down in the upper Co layer, respectively. The AS axis and S axis are along 123° and 33°, respectively. **f**, Azimuthal angular dependence of the switching field difference |Δμ₀H$_{SW}$| between the switching of AP+ to AP- and AP- to AP+. The maximum switching field difference $\Delta\mu_0 H_{SW}$ is 22.1 (±3.4) mT.

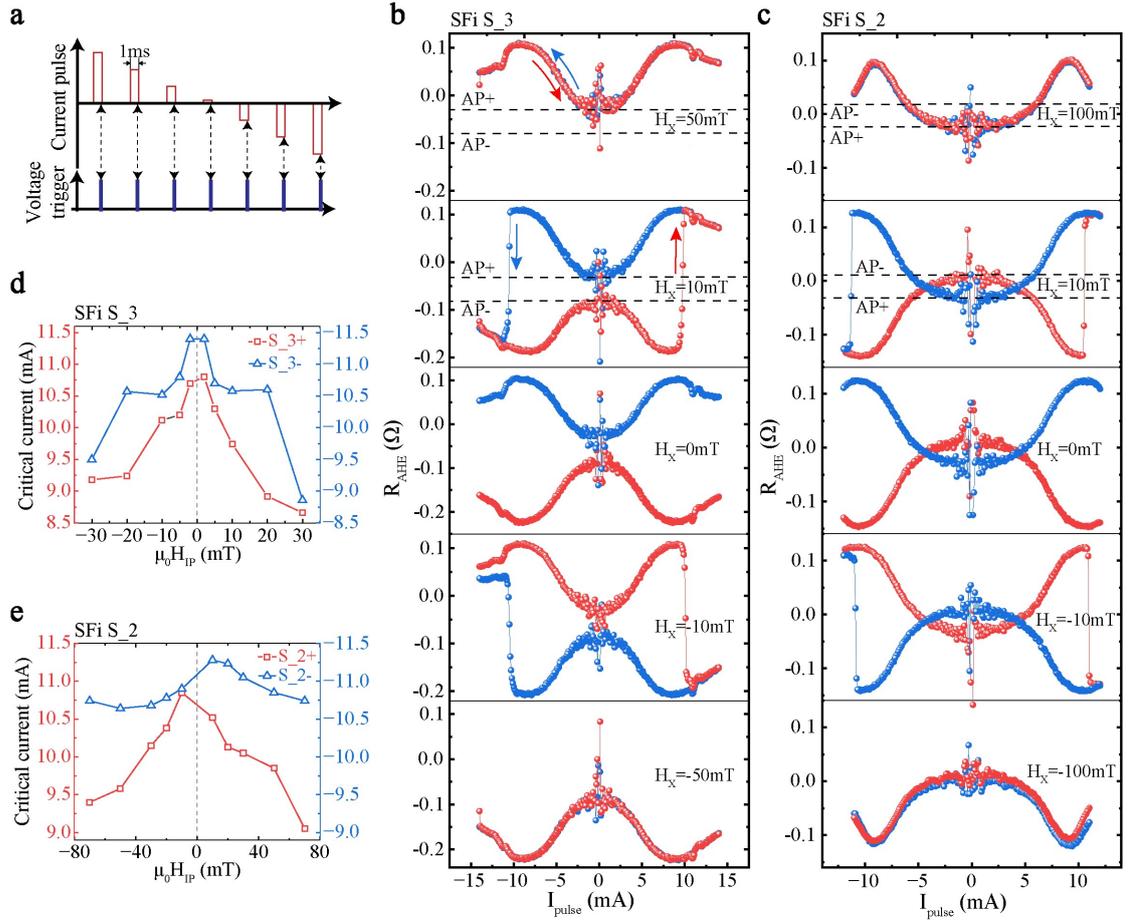

**Fig. 3 | Detection of the transient resistance states of SFi S_3 and S_2 during the SOT switching. a**, Schematic illustration of the transient state detection method for SOT switching. The AHE voltage detection occurs at the moment the applied SOT pulse ends. The pulse width of SOT is 1 ms. **b-c**, $R_{AHE}$ versus SOT pulse curves of S_3 (**b**) and S_2 (**c**) under different IP fields. Under the action of IL-DMI, SOT switching is no longer a dual-resistance state switching. The red and blue data represent the ascending and descending current scanning results of the SOT pulse, respectively. The horizontal dashed lines represent the resistance values of the two AP states of SFi. It should be noted that compared with S_3, the sign of the AP+ and AP- signal of S_2 is reversed due to the thicker bottom Co. No SOT switching occurs without $H_X$. With

proper $H_X$ assistance, deterministic SOT switching can be achieved. However, when $H_X$ exceeds a certain value (50 mT for S_3 and 100 mT for S_2), no SOT switching occurs and the curves of positive and negative current sweeping coincide. Due to the influence of IL-DMI, SOT current and thermal effects, the magnetization of the lower layer Co will gradually turn to IP with the increase of current. The upper and lower layers tend to align perpendicularly before switching. **d-e**, Asymmetric critical current distribution with different $H_{IP}$ of S_3 (**d**) and S_2 (**e**). During the SOT switching process of these two samples, regardless of the direction of $H_{IP}$, the critical current in the positive scan is always smaller than that in the negative scan.

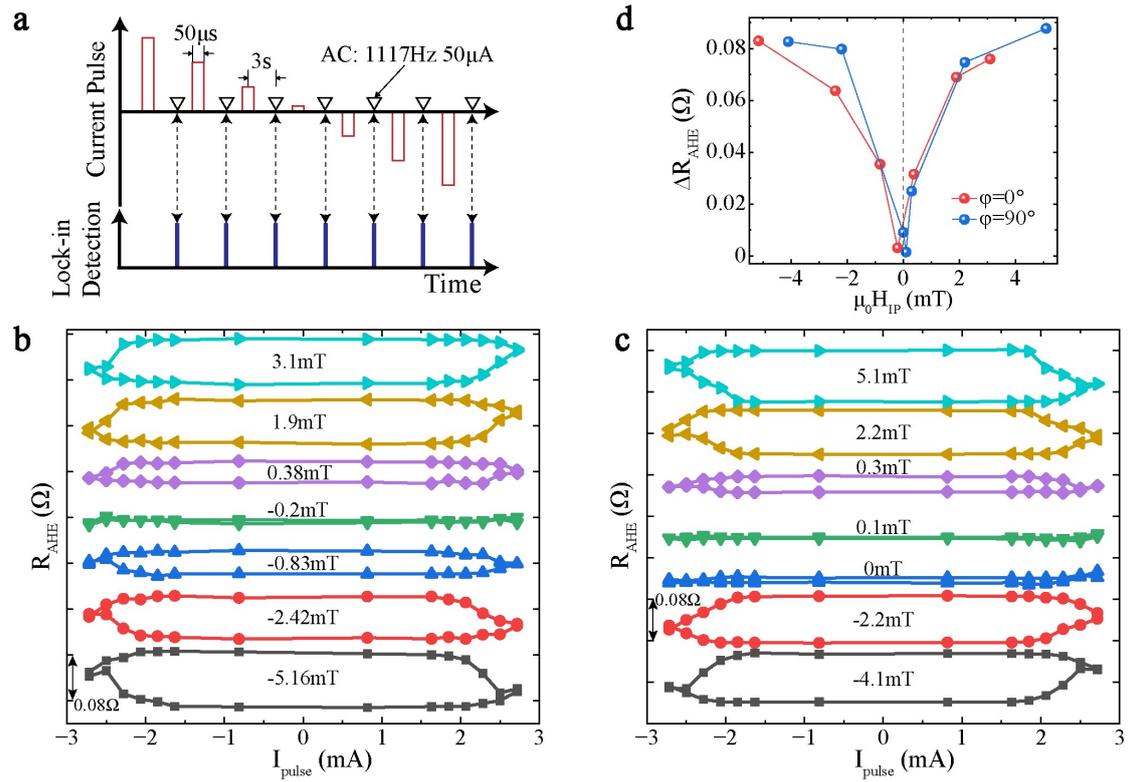

**Fig. 4 | Detection of the steady resistance states of SFi S_2 during the SOT switching. a**, Schematic illustration of the steady state detection method for SOT switching. AHE detection occurs 3 s after SOT pulse application, achieved with small AC current and lock-in technology. The pulse width of SOT is 50 μs. The amplitude of AC current is 50 μA and the frequency is 1117 Hz. **b-c**, $R_{AHE}$ versus SOT pulse curves of S_2 with Hall bar angle at φ=0° (**b**) and φ=90° (**c**) under different IP fields. The effective fields produced by IL-DMI at φ=0° and φ=90° are 0.2 mT and -0.1mT, respectively, as shown by the green curves. **d**, The dependence of the absolute value of the AHE resistance change with IP field at φ=0° and φ=90°.

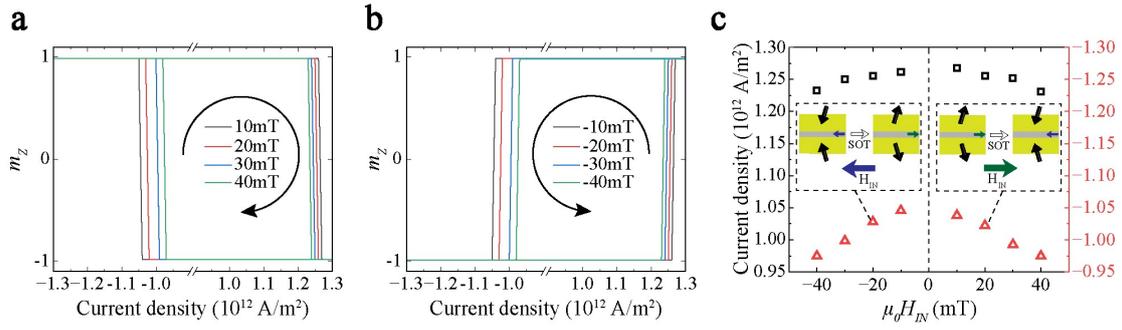

**Fig. 5 | Micromagnetic simulation of asymmetric critical switching current distribution. a-b**, Magnetization switching of the SFi versus SOT current density with the assistance of positive (**a**) and negative (**b**) IP fields. Switching the IP field direction will switch the SOT polarity. **c**, Asymmetric critical current distribution with different IP fields. Regardless of the direction of $H_{IP}$, the critical current in the negative scan (red triangle) is always smaller than that in the positive scan (black square). The inset explains that the reason for this asymmetry is because $H_{IN}$ and $H_{IL-DMI}$ always switch simultaneously.